# International comparison of optical frequencies with transportable optical lattice clocks

*International Clock and Oscillator Networking (ICON) Collaboration*[*]

**Optical clocks have improved their frequency stability and estimated accuracy[1,2,3,4,5,6,7] by more than two orders of magnitude over the best caesium microwave clocks that realise the SI second. Accordingly, an optical redefinition of the second has been widely discussed[8,9,10], prompting a need for the consistency of optical clocks to be verified worldwide[8,10]. While satellite frequency links are sufficient to compare microwave clocks, a suitable method for comparing high-performance optical clocks over intercontinental distances is missing. Furthermore, remote comparisons over frequency links face fractional uncertainties of a few $10^{-18}$ due to imprecise knowledge of each clock's relativistic redshift, which stems from uncertainty in the geopotential determined at each distant location. Here, we report a landmark campaign towards the era of optical clocks, where, for the first time, state-of-the-art transportable optical clocks from Japan and Europe are brought together to demonstrate international comparisons that require neither a high-performance frequency link nor information on the geopotential difference between remote sites. Conversely, the reproducibility of the clocks after being transported between countries was sufficient to determine geopotential height offsets at the level of 4 cm. Our campaign paves the way for redefining the SI second and has a significant impact on various applications, including tests of general relativity[11], geodetic sensing for geosciences, precise navigation, and future timing networks.**

[*] A list of authors and their affiliations appears at the end of the paper.

Confirmation of the consistency of optical clocks, especially between those developed by different institutes, is an essential step toward redefining the SI second[10]. To achieve frequency comparisons that match the performance of optical clocks, optical fibre links[12] and free-space optical links[13] have been developed with instabilities even lower than those of the optical clocks. While optical fibre links have been successfully established connecting metrological laboratories within Europe[12], the US[13], and Japan[14], similar links on an intercontinental scale are yet to be demonstrated. A complementary approach involves measuring the frequency ratios of different clock transitions locally and comparing the values internationally[13,15], providing an indirect assessment of clock reproducibility on a global scale.

A transportable optical clock enables worldwide on-site comparisons to assess the consistency of clock frequencies. Beyond frequency metrology, such a clock is also anticipated to have various other applications. The exceptional performance of optical lattice clocks, achieving fractional instability at $10^{-18}$ in short averaging times, combined with high-performance links[12,13], allows for precise measurements of relativistic time dilation experienced by clocks differing in geodetic elevation at the centimetre level. A transportable optical lattice clock using strontium ($^{87}$Sr) atoms was developed by a group at PTB (Physikalisch-Technische Bundesanstalt)[16]. They demonstrated its field operation and geodetic measurement by connecting their clock to a reference clock at INRiM (Istituto Nazionale di Ricerca Metrologica)[17]. A group at RIKEN has also developed a transportable system[18] and demonstrated an experiment to verify general relativity in a broadcast tower[11]. The field operation of highly accurate and stable optical lattice clocks is particularly promising for applications such as chronometric levelling, seismology, and volcanology[19].

We demonstrate an international comparison of strontium-based optical lattice clocks developed independently at NPL (National Physical Laboratory, United Kingdom), PTB (Germany) and RIKEN (Japan) employing transportable lattice clocks and optical fibre links. These comparisons make an important contribution towards a redefinition of the SI second[10], serve as demonstration of chronometric levelling, and demonstrate a high resolution non-local clock comparison that does not require a frequency link or knowledge of the redshift arising from the geopotential difference between distant sites.

The first phase of the international comparison campaign took place at NPL in early 2023. A transportable clock from RIKEN[18] was shipped by air from Japan to the UK. A transportable clock from PTB, which is an upgraded version of Ref. 16, was housed in a trailer and delivered overland from Germany to the UK. The transportable clocks were compared locally with the stationary clock at NPL and remotely with the stationary clock at PTB via a 2155-km-long European optical fibre link[12] (EFL), as shown in Fig. 1. The transportable clocks from RIKEN and PTB were then brought overland to PTB and compared locally with the stationary clock at PTB for the second phase of the campaign.

## Experiment

The frequencies of the involved $^{87}$Sr lattice clocks were compared locally through direct optical beat note measurements. Optical frequency combs facilitated the connection of the clocks to the EFL for remote comparison and enabled the local transfer of ultrastable light originating at other wavelengths. All optical links were equipped with phase noise cancellation.

At each site, the PTB transportable clock (Fig. 1a, denoted PTB(T)) was situated in a trailer parked outside the laboratory building, while its clock laser was stabilized to a rigidly mounted transportable cavity[20] inside the laboratory. Atoms were probed in a copper enclosure maintained at 223 K to reduce the uncertainty of the blackbody radiation (BBR) shift. The RIKEN transportable clock[11,18], consisting of two half-height 19” rack laser units and a physics package (Fig. 1b), was housed inside a laboratory at each site, and included a BBR shield, with its temperature controlled at 245 K by a Peltier cooler. The RIKEN and, optionally, the PTB(T) transportable clocks were probed by a laser phase-locked to the clock laser from the NPL (PTB) stationary clock.

In the NPL stationary clock[21], the clock transition was probed in a room-temperature environment, with temperature precisely monitored by thermometers around the vacuum chamber. The clock laser was stabilized to a 48.5-cm-long room-temperature ultra-low expansion glass (ULE) cavity[12], with fractional frequency instability of $7 \times 10^{-17}$ at intermediate timescales. In the PTB stationary clock, the clock transition was probed inside a room-temperature BBR shield, with the clock laser stabilized to either a cryogenic silicon cavity[22] or a 48-cm-long room-temperature ULE cavity[23], with fractional frequency instabilities of $4 \times 10^{-17}$ and $8 \times 10^{-17}$, respectively, at intermediate time scales.

The estimated systematic uncertainties of the clocks' frequencies are about $6 \times 10^{-18}$ and $1.4 \times 10^{-17}$ for the RIKEN and PTB(T) transportable clocks and $3 \times 10^{-18}$ and $1.4 \times 10^{-17}$ for the PTB and NPL stationary clocks, respectively.

We determine the differential relativistic redshifts between the clocks using local geometric levelling with respect to the height reference markers G4L10 at NPL or PB02 at PTB[24], which are close to the respective stationary strontium clocks, in combination with the well-known local gravity accelerations[24] at each site. Locally, these measurements are sufficient to determine the shift with negligible uncertainty. To compare clocks located at different sites via the EFL, we additionally need to determine the geopotential difference, $\Delta W$, between the reference markers to correct for the relativistic redshift. We follow the approach of previous publications[24,25] and use the result of the well-established GNSS/geoid approach (GA), $\Delta W^{\mathrm{GA}} = 667.30(29)\ \mathrm{m^2\ s^{-2}}$ from G4L10 to PB02,[24] for this. To demonstrate a complementary approach, we also determine the geopotential difference from the measurements presented here chronometrically and use this result to correct for the redshift between the NPL and PTB stationary clocks.  Finally, we demonstrate a remote comparison of the NPL and PTB stationary clocks via the transportable clocks, which does not require knowledge of the geopotential difference $\Delta W$.

## Results of clock comparisons

Time traces of the measured frequency ratios between the clocks, including corrections for systematic shifts and the relativistic redshifts are shown in Fig. 2a. In phase one of the campaign, the frequency ratios $R_{ij} = \nu_i / \nu_j$ with $i, j$ = NPL, PTB, PTB(T), and RIKEN were recorded. In the second phase, the NPL clock could not contribute due to interruptions in the comparator chain. The instabilities for the frequency ratios $R_{ij}$ are shown in Fig. 2b. We observe excellent instabilities that in the best cases enabled a fractional statistical resolution of $1 \times 10^{-18}$ to be achieved in just 3 hours of averaging time. The best instabilities were achieved for the RIKEN/NPL comparison, where synchronous operation of both clocks cancelled common-mode noise of the sampled clock laser[26] that would otherwise reduce stability via the Dick effect[27], and for the RIKEN/PTB comparison at PTB, where the superior stability of the cryogenic silicon cavity[22] allowed for comparable results in asynchronous operation of the clocks. The instability of the NPL/PTB comparison is higher than that of the other two due to additional phase noise from the EFL at short averaging times and the absence of common-mode rejection of clock laser noise between the clocks.

In phase one, comparison between four clocks yields twelve possible ratios $R_{ij}$ with an expected value of 1. The deviations of the six unique ratios from 1 are plotted by filled symbols in Fig. 3. The remaining six ratios, with the denominator and numerator inverted, are opposite signs of the plotted values. In phase two, three clocks were compared, yielding six ratio values. The three unique ratios are shown by open symbols in Fig. 3. All ratios are shown with error bars given by the total uncertainty $\sigma_{ij}$ from the quadrature sum of the statistical uncertainty of the measured ratio and the systematic uncertainty of each clock $i$ and $j$, and for remote comparisons via the EFL an uncertainty contribution of $3.3 \times 10^{-18}$ from $\Delta W^{\mathrm{GA}}$.

The ratios in Fig. 3 reveal the level of reproducibility from the transportable clocks after they have been moved between countries. As shown by green circles, we observed good agreement between the same clock ratios that were measured at NPL and then again at PTB. This demonstrates the reliability of these transportable clocks to provide a measurement connection between distant stationary systems. Therefore, we can also measure the frequency ratio $R_{\mathrm{NPL,PTB}}$ of the stationary optical clocks at NPL and PTB without precise knowledge of the geopotential or a high-performance frequency link between both sites. Instead, it is computed from the frequency ratios between the respective stationary clock at NPL and PTB and either transportable clock as $R^{(x)}_{\mathrm{NPL,PTB}} = R_{\mathrm{NPL},x}/R_{x,\mathrm{PTB}}$ where $x$ = RIKEN, PTB(T), requiring only local relativistic redshift differences to be accounted for. We determine a weighted average using the data from both transportable clocks and regard their systematic effects as correlated between the measurements at PTB and NPL. This indirect measurement of the NPL/PTB frequency ratio achieves the same uncertainty as the measurement via the EFL, as shown in Fig. 3.

In order to compare our measurement to other clock comparisons, we introduce the extended weighted root mean square deviation (WRMSD) as a new figure of merit. In our case, the extended WRMSD (EWRMSD) for a level of confidence $p$ is given by

$$t_p(\nu)\sqrt{\frac{\sum_{i,j\neq i} w_{ij}\left(R_{ij}-1\right)^2}{\sum_{i,j\neq i} w_{ij}}}$$

i.e., the RMS deviation of the frequency ratios from unity with weights $w_{ij} = 1/\sigma_{ij}^2$ and an effective coverage factor[28] $t_p(\nu)$ that depends on the measurement's number of degrees of freedom, $\nu$, and the confidence level (see Methods for details). It quantifies the level of agreement of a measurement including the higher confidence gained by comparing a larger number of clocks, without relying heavily on the estimated measurement uncertainties. Following usual metrological procedure, we consider a confidence level $p \approx 0.9545$ (denoted as $k_p = 2$) in the following.

The deviations of the frequency ratios from unity that we observe are somewhat larger than expected from the measurement uncertainties. We find EWRMSD values of $1.0 \times 10^{-16}$ for phase one of the campaign and $9 \times 10^{-17}$ for phase two, with effective coverage factors of $t_p(\nu = 3) = 3.31$ and $t_p(\nu = 2) = 4.53$, respectively.

We have computed the EWRMSD of previous clock comparisons (see Methods for the extension to comparisons between clocks based on different species or transitions) and compare them to our results in Fig. 4. Our measurements demonstrate a high level of agreement compared to most previous international or interlaboratory clock comparisons, ranking among the highest for clocks of the same kind. Comparisons between different institutes provide a higher degree of independence and are thus required in the roadmap for the redefinition of the second[10]. Our measurement makes an important contribution to this task.

**Chronometric levelling**

Beyond the metrological aspects, we can use our comparisons of the transportable clocks with a stationary clock, both locally and remotely via the EFL, for chronometric levelling (CL). Here, we require the transportable clocks' frequencies to be reproducible after transportation, which is corroborated by the comparisons between the transportable clocks shown in Fig. 3. To derive the geopotential difference between the reference markers at NPL and PTB, we only correct for the local height differences of the clocks to the respective reference marker at each site. For each transportable clock, the chronometrically measured geopotential difference is given by

$$\Delta W_i^{\mathrm{CL}} = c^2 \left( R_{i,\mathrm{PTB}}^{\mathrm{rem}} - R_{i,\mathrm{PTB}}^{\mathrm{loc}} \right)$$

with $c$ the speed of light, $R_{i,\mathrm{PTB}}^{\mathrm{rem}}$ and $R_{i,\mathrm{PTB}}^{\mathrm{loc}}$ the remote and local frequency ratio with the PTB stationary clock, respectively, and $i$ = RIKEN, PTB(T). We measure $\Delta W_{\mathrm{RIKEN}}^{\mathrm{CL}}$ = 668.18(58) $\mathrm{m^2\,s^{-2}}$ and $\Delta W_{\mathrm{PTB(T)}}^{\mathrm{CL}}$ = 667.87(49) $\mathrm{m^2\,s^{-2}}$. The respective uncertainties result mainly from the statistics of the measurement; systematic effects are regarded as fully correlated. The weighted average of the two measurements is $\overline{\Delta W}^{\mathrm{CL}}$ = 668.00(37) $\mathrm{m}^2\,\mathrm{s}^{-2}$. Its uncertainty corresponds to less than 4 cm in height and is comparable to the uncertainty of 0.29 $\mathrm{m}^2\,\mathrm{s}^{-2}$ resulting from the best established geodetic methods.[36]

Figure 5 shows a comparison of our chronometric result to previous determinations by geometric levelling[29,30,31,32,33] within the European Vertical Reference Frame (EVRF) and by the GNSS/geoid approach[34,35,36] via the European Gravimetric (Quasi) Geoid (EGG). The EGG values have become more accurate with the use of dedicated satellite gravity missions (CHAMP, GRACE, GOCE). The EVRF2007 (ITOC) and EVRF2019 values have changed significantly towards the EGG values by recent analysis of a levelling measurement through the Channel Tunnel and partially by considering a tilt[33] of the old French levelling network IGN69. We find agreement between our result and the GNSS/geoid approach. By our measurement, we have established an independent and direct connection between the British and German height reference systems. Moreover, the chronometric levelling measurement was conducted in a two-month measurement campaign, which can be readily repeated, while the geometric levelling approach is based on historic data that cannot be reproduced as easily[33].

## Conclusion

We have conducted an international comparison of four same-species optical clocks built and operated by three independent institutes spanning two continents. The comparison, which was carried out in two phases, was enabled by the transportability of two of the clocks (from RIKEN and PTB), which were hosted together first at NPL, and later at PTB. The campaign benefited from the use of the EFL which allowed chronometric levelling between NPL and PTB when used together with the stationary and transportable clocks.

We introduced the extended WRMSD as a new figure of merit for clock agreement and compared our results to previous measurements. Our comparisons achieved some of the highest levels of agreement to date for clocks developed by independent institutes, reaching an extended WRMSD of $9 \times 10^{-17}$ at 95.45 % confidence in the best case. This level of agreement is the best achieved among independent clocks of the same kind and well below the expanded uncertainty of the best caesium primary frequency standards, which makes it an important step towards a redefinition of the SI second.

Notably, our measurements are the first to demonstrate two capabilities that will be important for frequency metrology in the future: First, the ability to compare optical clocks at remote locations with an uncertainty below that of primary standards without requiring a direct link or precise knowledge of the geopotential difference. This allows for comparisons of optical clocks across intercontinental distances at the limit of their uncertainties. Second, the chronometric determination of the potential difference between locations across large distances with an uncertainty that is on par with the best of the established geodetic methods. Given the limitations of the latter at a level of $3 \times 10^{-18}$, chronometric levelling will become indispensable for comparing optical clocks over large distances, even at the present state of the art of around $1 \times 10^{-18}$ uncertainty for optical clocks.

In the future, transportable clocks will become crucial in investigating the consistency of frequency metrology at the $10^{-18}$ level on an intercontinental scale. Such highly reproducible and stable transportable clocks, as demonstrated in this work, will open up new possibilities for a wide range of applications, not only in time and frequency metrology, but also in high-resolution chronometric levelling[19] and tests of fundamental physics[37].

**International Clock and Oscillator Networking (ICON) Collaboration***

Anne Amy-Klein[1], Erik Benkler[2], Pascal Blondé[3], Kai Bongs[4], Etienne Cantin[1], Christian Chardonnet[1], Heiner Denker[5], Sören Dörscher[2], Chen-Hao Feng[6], Jacques-Olivier Gaudron[6], Patrick Gill[6], Ian R Hill[6†], Wei Huang[6], Matthew Y H Johnson[6], Yogeshwar B Kale[7], Hidetoshi Katori[8,9,10‡], Joshua Klose[2], Jochen Kronjäger[2], Alexander Kuhl[2], Rodolphe Le Targat[3], Christian Lisdat[2§], Olivier Lopez[1], Tim Lücke[2], Maxime Mazouth[3], Shambo Mukherjee[2], Ingo Nosske[2], Benjamin Pointard[3], Paul-Eric Pottie[3], Marco Schioppo[6], Yeshpal Singh[7**], Kilian Stahl[2], Masao Takamoto[9,10], Mads Tønnes[3], Jacob Tunesi[6], Ichiro Ushijima[8,9], Chetan Vishwakarma[2]

1 Laboratoire de Physique des Lasers (LPL), Université Sorbonne Paris Nord, CNRS, 99 Avenue Jean-Baptiste Clément, 93430 Villetaneuse, France.

2 Physikalisch-Technische Bundesanstalt, Bundesallee 100, 38116 Braunschweig, Germany.

3 LNE-SYRTE, Observatoire de Paris, PSL Research University, CNRS, Sorbonne Universités, UPMC University of Paris 06, 61 Avenue de l'Observatoire, 75014 Paris, France.

4 Deutsches Zentrum für Luft- und Raumfahrt (DLR), Institut für Quantentechnologien, Wilhelm-Runge-Straße 10, 89081 Ulm, Germany.

5 Institut für Erdmessung, Leibniz Universität Hannover, Schneiderberg 50, 30167 Hannover, Germany.

6 National Physical Laboratory, Hampton Road, Teddington TW11 0LW, United Kingdom.

7 School of Physics and Astronomy, University of Birmingham, Birmingham B15 2TT, United Kingdom.

8 Department of Applied Physics, Graduate School of Engineering, The University of Tokyo, Bunkyo-ku, Tokyo 113-8656, Japan.

9 Quantum Metrology Laboratory, RIKEN, Wako, Saitama 351-0198, Japan.

10 Space-Time Engineering Research Team, RIKEN Center for Advanced Photonics, Wako, Saitama 351-0198, Japan.

† ian.hill@npl.co.uk

‡ hkatori@riken.jp, katori@amo.t.u-tokyo.ac.jp

§ christian.lisdat@ptb.de

** y.singh.1@bham.ac.uk

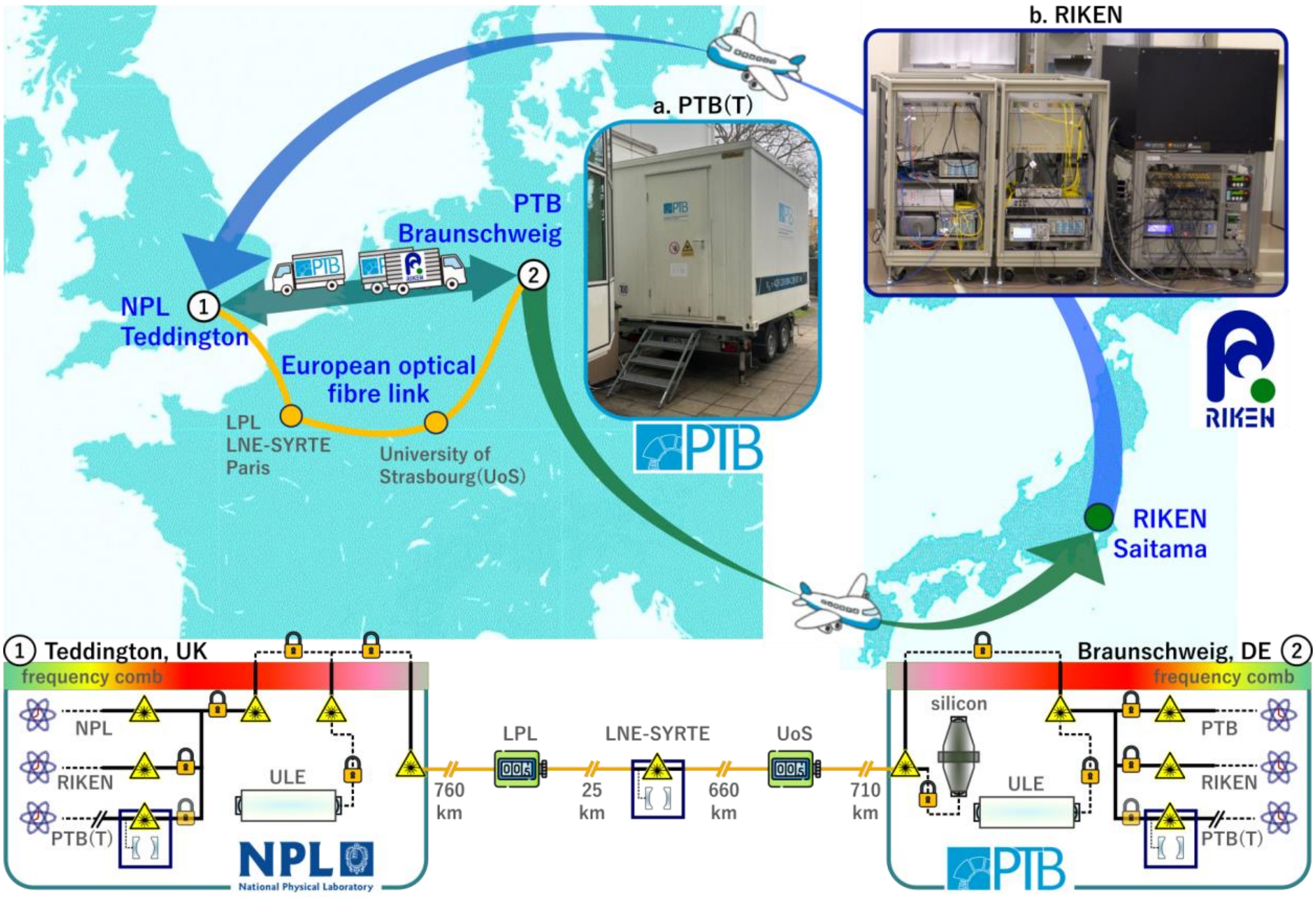

**Figure 1** International comparison of optical lattice clocks. The campaign consists of four Sr-based optical lattice clocks: two transportable clocks from RIKEN (Japan) and PTB (Germany), and two stationary clocks at NPL (UK) and PTB connected by the EFL. The campaign was held between the beginning of March and end of April 2023, including time for preparation and transportation between the sites.

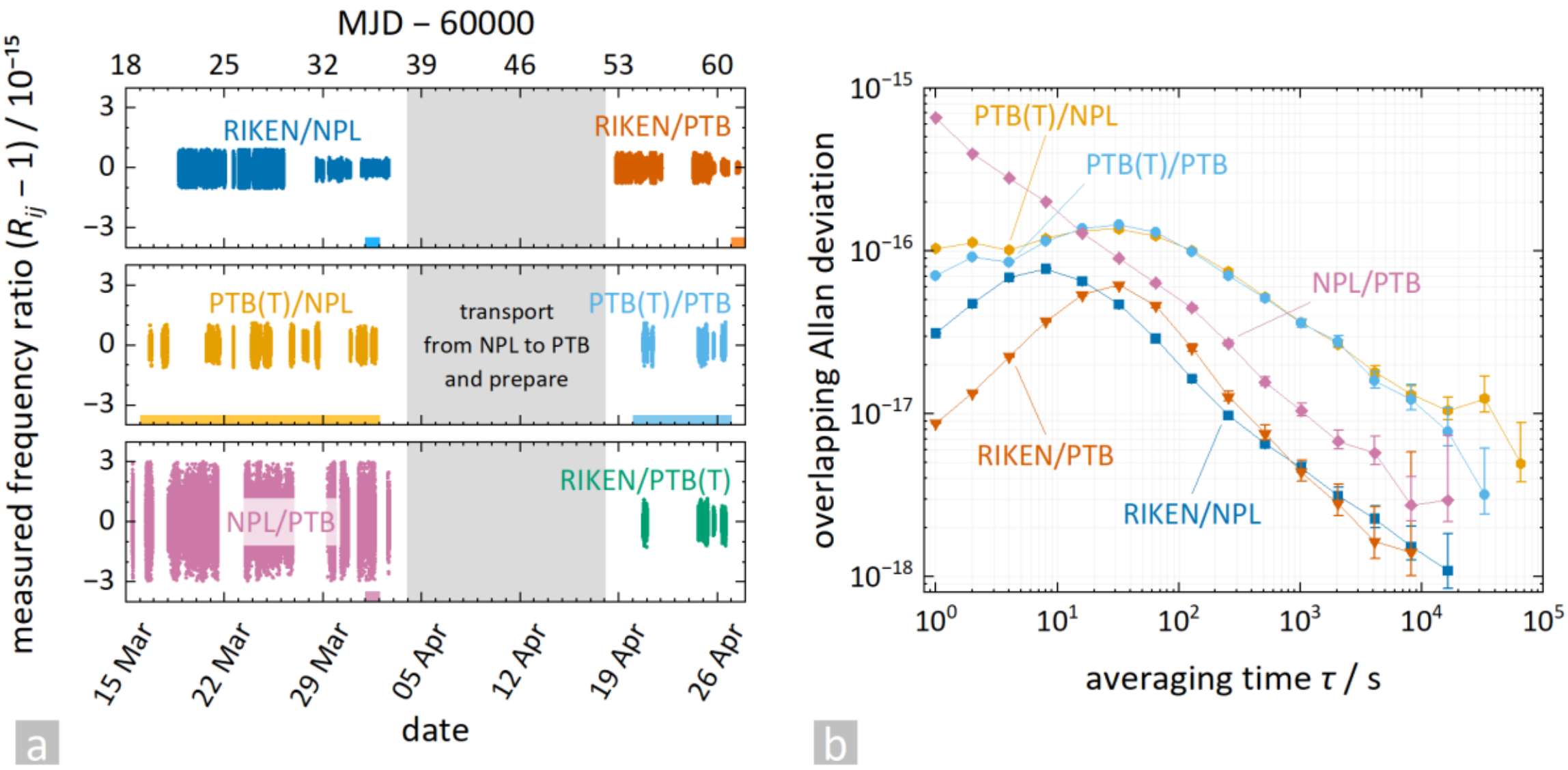

**Figure 2 a**. Frequency ratios $R_{ij}$ of the strontium clocks. Selected ratios are shown to indicate the uptime of each clock. The frequency ratio NPL/PTB shown here uses the relativistic redshift determined using the GNSS/geoid approach. The bars below the traces indicate the periods used to compute the Allan deviations shown in panel b. **b**. Instability of selected frequency ratios $R_{ij}$. The overlapping Allan deviation has been computed for either a single day of high stability operation or the entire trace, as indicated by the bars in panel a. Days with synchronous interrogation of the NPL and RIKEN clocks and using the ultrastable silicon resonator at PTB have been chosen, respectively. The error bars indicate the 68.3 % confidence intervals assuming white frequency noise.

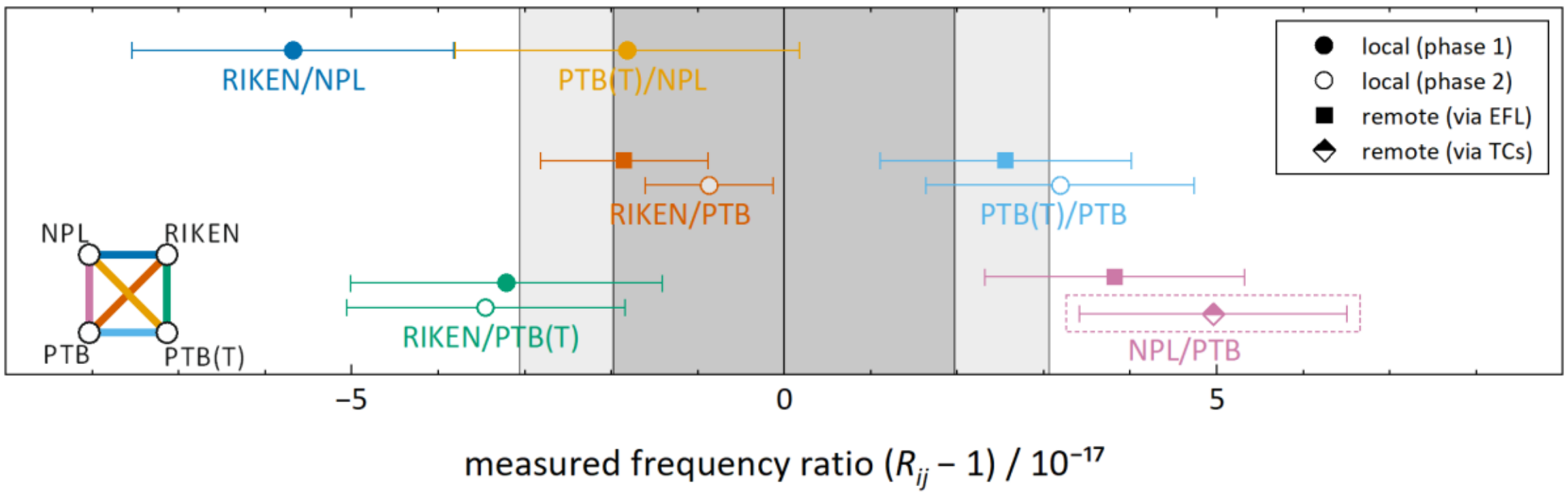

**Figure 3** Mean frequency ratios $R_{ij}$ measured during the two phases of the measurement campaign. Colours indicate the clock pairs as shown in the inset. The WRMSD of the ratios during phase one (two) is shown as a light (dark) shaded region. The NPL/PTB frequency ratio measured indirectly by comparison to the transportable clocks (TCs) is highlighted by a dashed box.

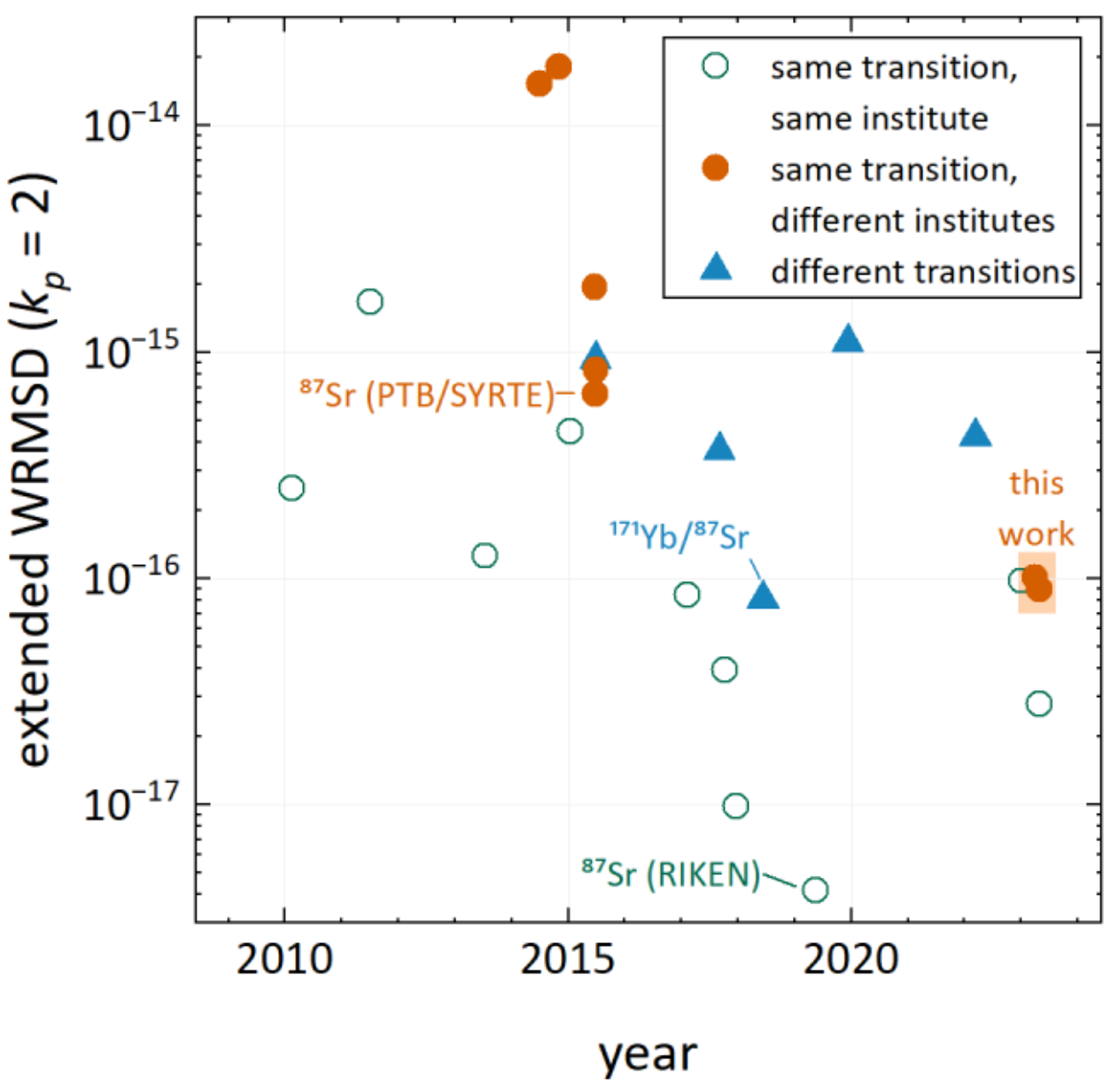

**Figure 4** Agreement of optical clock comparisons between different institutes (filled symbols) and within institutes (open symbols), as represented by the extended WRMSD ($k_p$ = 2). Circles (triangles) denote comparisons of clocks using the same (different) reference transitions. The shaded box indicates the results of this measurement, which are stated in the main text. See Extended Data Table 1 for further details and references.

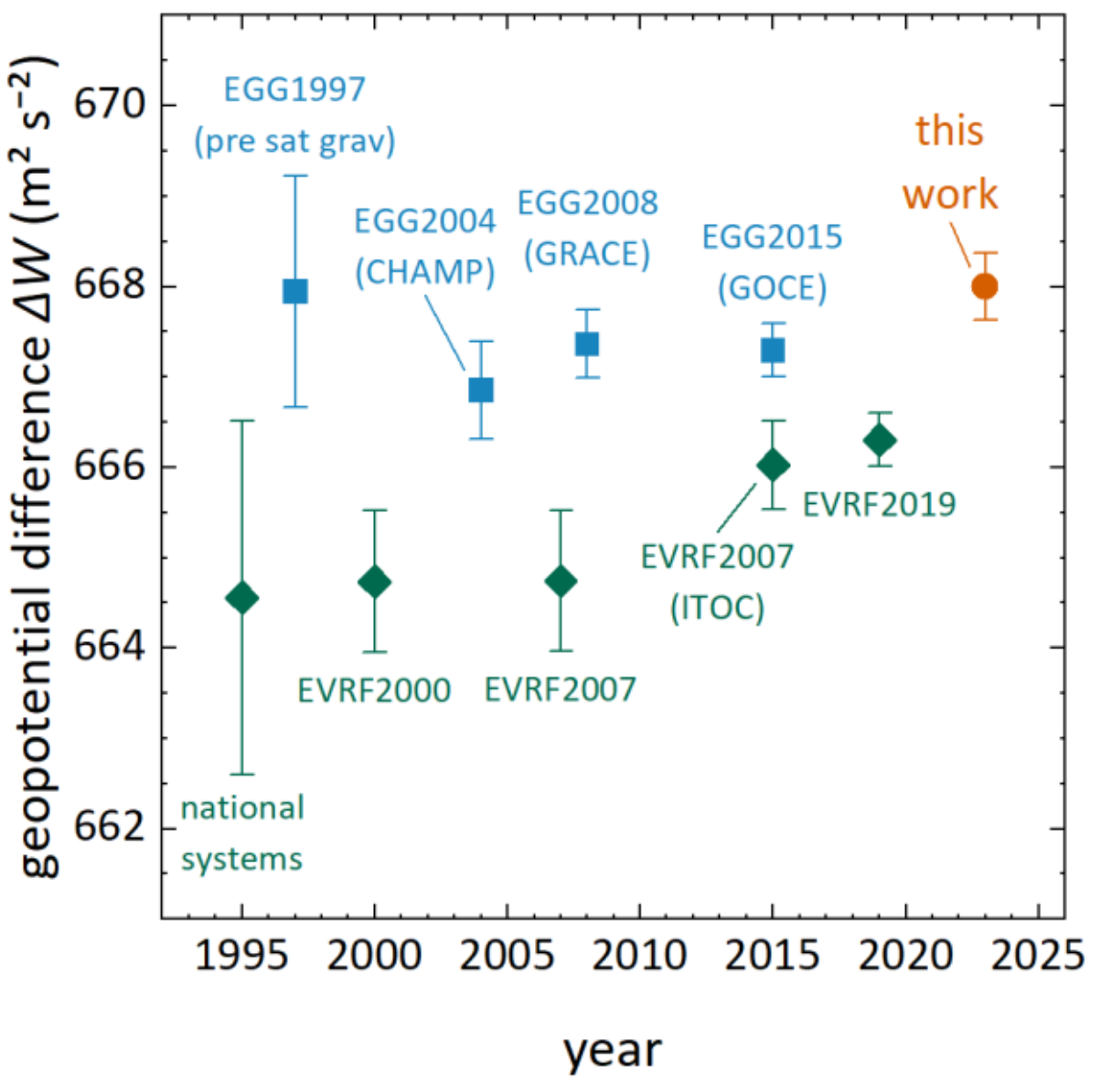

**Figure 5** Determinations of the geopotential difference between PTB and NPL over time. The geopotential difference $\Delta W$ between the reference markers PB02 at PTB and G4L10 at NPL, as determined by geometric levelling (green diamonds), the GNSS/geoid approach (blue squares), and chronometric levelling (red circle) is shown.

# Methods

## Description of individual clocks and their operation

### *RIKEN transportable clock*

Details of the transportable clock from RIKEN can be found in Ref. 11, and details of the laser system in Ref. 18. The system is equipped with a blackbody radiation (BBR) shield to reduce the BBR shift uncertainty. The temperature of the shield is stabilized by a Peltier cooler at 245 K, reducing the BBR shift to $-2316.6 \times 10^{-18}$ and its uncertainty to $2.6 \times 10^{-18}$, given mainly by the uncertainty of the thermal sensor.

The optical lattice is formed in a bow-tie optical cavity to create a well-defined standing wave, allowing atoms to move from the lattice loading region into the BBR shield. The clock was operated in an operational magic intensity condition[38], where the intensity dependence of the lattice light shift is minimized, at a trap depth of 70(3) $E_r$ (where $E_r$ is the lattice photon recoil energy) and an average axial vibrational state population of atoms $n = 0.1(2)$.

The clock transition was interrogated by Ramsey spectroscopy with a $\pi/2$ pulse duration of 17 ms or 29 ms and a dark time of 300 ms or 400 ms. The uncertainty of the probe light shift is evaluated to be $1.7 \times 10^{-18}$, which is attributed to the uncertainties of the clock laser intensity and the probe light shift coefficient[11]. The intensity was evaluated with an uncertainty of 2.0 % by analyzing the observed Ramsey spectrum.

Interactions between atoms in the lattice cause the density shift. Assuming that the density shift is proportional to the number of trapped atoms, we evaluated the clock frequency shift at various atom numbers and obtained the density shift coefficient of −1.88(30) μHz/atom for a typical number of atoms $N \sim 2{,}000$

***PTB transportable clock***

The transportable clock from PTB is housed in an air-conditioned car trailer. $^{87}$Sr atoms are loaded into a vertical optical lattice, derived from a titanium:sapphire laser, and transported into a BBR shield for clock interrogation[11]. The atoms are spin-polarized to the $m_F = \pm 9/2$ magnetic sublevels. We then perform Rabi spectroscopy on the $^3P_0$–$^1S_0$ transition with a pulse length of typically 350 ms, which leads to a probe light shift and uncertainty of $<1 \times 10^{-18}$. One total sequence length is typically about 2 s. Light to drive the clock transition is generated by a diode laser operating at the sub-harmonic of the clock transition and stabilized to a high-finesse optical resonator[20].

For the lattice light shift determination, the clock was operated in an interleaved mode switching between the two trap depths 31(3) $E_r$ and 51(5) $E_r$. The fraction of atoms with axial states $n_z > 0$ was typically around 0.5(1). The estimated lattice light shift uncertainty at the different parts of the campaign was between $1.0 \times 10^{-17}$ and $2.0 \times 10^{-17}$ and dominated by available statistics.

This also represents the total systematic uncertainty of the clock during each part of the measurement campaign, as all other uncertainty contributions are estimated to be below $2.0 \times 10^{-18}$: The density shift was characterized by interleaving between two cycles with different atom numbers. Careful characterization of the BBR shield geometry and an operating temperature of 223 K results in a BBR shift uncertainty below $1 \times 10^{-18}$.

***NPL stationary clock***

A description of the NPL stationary clock NPL-Sr1 and its frequency evaluation is given in Ref. 21. For the reported measurements, changes include an out-of-vacuum lattice enhancement cavity with a finesse of 70 and waist of 153 µm. Operating with an atom number close to $N \sim 1{,}000$, we resolve no density related shifts at the level of $8 \times 10^{-19}$, evaluated through interleaved measurements with modulated atom number up to $N \sim 4{,}000$.

We use an updated coefficient for the background gas collisional shift[39] of $-3.0(3) \times 10^{-17}$ / ($\tau$/s), where $\tau$ is the vacuum-limited 1/$e$ trap lifetime and confirm hydrogen as dominant using a residual gas analyzer. We arrive at a correction of $6.4(10) \times 10^{-18}$ based on a lattice trap lifetime of 4.7 s. The optical lattice was operated at a depth of 70 $E_r$ and the light shift evaluated as in Ref. 21 arriving at a correction of $-5.0(22) \times 10^{-18}$.

The BBR shift is evaluated with updated coefficients[40] and confirmed by in-situ thermometry carried out after the measurement campaign using a precisely calibrated Pt100 in-vacuum sensor located at the position of the atoms. No updated corrections were applied using this method, though the shift estimated by the external ensemble method described in Ref. 21 and used here was shown to agree within its $1\sigma$ uncertainty. The quadratic Zeeman shift was assessed dynamically with a representative mean correction of $233.7(3) \times 10^{-18}$. The uncertainty is dominated by the correction coefficient[41] with negligible contribution from spread in the measured line splitting.

The clock was operated using both Rabi spectroscopy with $\pi$ pulse duration between 320 ms and 490 ms, and Ramsey spectroscopy with a $\pi/2$ pulse duration of 17 ms and a dark time between 300 ms and 400 ms. The probe Stark shift associated with Ramsey spectroscopy was determined as $0(2) \times 10^{-18}$, measured in interleaved comparison to a 400 ms Rabi probe, for which we assign a correction of $0(1) \times 10^{-18}$.

***PTB stationary clock***

The PTB-Sr3 laboratory clock uses a temperature-monitored heat shield, which was operated at room temperature, to control the BBR shift at the level of $1 \times 10^{-18}$. Viewports provide optical access; an atomic beam is injected from an external atom source through free-space access ports. Its vertical optical lattice was operated at different lattice depths between 75 $E_r$ and 150 $E_r$, allowing us to confirm the stability of the lattice light shift at the level of $3 \times 10^{-19}$ by a simultaneous characterization measurement.

A similar procedure as described in previous publications[42,43] was used for state preparation, interrogation, and read-out, but we additionally employed sideband cooling on the $^3P_1$–$^1S_0$ transition to transfer >90 % of the atoms into the longitudinal vibrational ground state. The $^3P_0$–$^1S_0$ clock transition was probed using Rabi spectroscopy with a $\pi$ pulse duration of about 1 s, where the atoms initially occupied the excited state. Feedback was applied directly to the frequency of the 429 THz clock laser, which was pre-stabilised to the ultrastable ULE cavity in the early stages of the second measurement phase and to the cryogenic silicon cavity otherwise. Evaluation of the systematic frequency shifts follows the description in Ref. 44.

**Evaluation of measurement uncertainty**

We calculate the uncertainty $\sigma_{ij}$ of each frequency ratio measurement between two clocks $i$ and $j$ by adding the statistical uncertainty, the systematic uncertainties of the clocks, and the uncertainty of the differential relativistic redshift between them in quadrature, separately for each part of the campaign.

The statistical uncertainty is estimated in two steps to check and, if necessary, account for excess scatter that is not compatible with white frequency noise: First, we compute the overlapping Allan deviation of the entire data set, determine the instability by least-squares fitting the function $\sigma_y(\tau) = \sigma_y^{(0)}\ (\tau/\mathrm{s})^{-1/2}$ for $\tau \geq \tau_0$ = 128 s and extrapolate this function to the full length of the data set. Second, we split the data set into a series of one-day intervals, determine their mean value and statistical uncertainty using the procedure from the first step, and compute the Birge ratio, $R_B$, of the series to check for excess scatter.[45] Similar to the procedure used in other clock comparisons[13], we inflate the uncertainty to account for possible excess scatter by multiplying the uncertainty determined in the first step by the larger of one or $R_B$ and use the result as statistical uncertainty.

The systematic uncertainties are determined according to the procedures established for each clock, neglecting potential, but likely small correlations between the clocks. Lastly, we neglect the uncertainties of local differential relativistic redshifts, which are at the level of $10^{-19}$, but include an uncertainty contribution from the geopotential difference between the height reference markers at NPL and PTB.

**Extended weighted root mean square deviation**

As a figure of merit of a set of comparisons between clocks that realize the frequencies of two transitions, *A* and *B*, we introduce the extended weighted root mean square deviation (EWRMSD)

$$S_p = t_p(\nu)\, S$$

for a given level of confidence $p$ and coverage factor[28] $k_p$. Here, the regular weighted root mean square deviation is given by

$$S = \sqrt{\frac{\sum_{i,j\neq i} w_{ij}\left(R_{ij} - y_{ij}\right)^2}{\sum_{i,j\neq i} w_{ij}}}$$

The weights $w_{ij} = \sigma_{ij}^{-2}$ are derived from the measurement uncertainties $\sigma_{ij}$. For clocks of the same kind ($A = B$), $y_{ij} = 1$ since the true frequency ratio is known to be unity. Otherwise ($A \neq B$), we use the average frequency ratio of the data set itself as an estimate of $y_{ij}$ as the true frequency ratio is not known. The effective coverage factor $t_p(\nu) \geq k_p$ accounts for the finite number of degrees of freedom, $\nu$, using the $t$-distribution, where $k_p$ is the coverage factor for the level confidence $p$ and $\lim_{\nu\to\infty} t_p(\nu) = k_p$.[28] We estimate $\nu$ to be the number of participating clocks minus one for clocks of the same kind, because any additional frequency ratio can be derived from the others, and the number of ratios measured minus one otherwise, because of the estimation of the true frequency ratio. In the main text, we use a level of confidence $p \approx 0.9545$ ($k_p = 2$).

The EWRMSDs of previous publications shown in Fig. 4 are summarized together with additional data in Extended Data Table 1 and have been computed from the references listed therein. The dates listed in Extended Data Table 1 are derived from the end of the respective measurement, if reported, and the date of publication otherwise. In case of multiple measurements or publications, the most recent date is used.

**Comparison of the stationary clocks using chronometric levelling**

Using the geopotential differences stated in the main text and shown in Fig. 5, the NPL/PTB frequency ratio measured via the EFL in phase one using the chronometric levelling result is readily derived from the result using the GNSS/geoid approach shown in Fig. 3 and the difference of the respective geopotential values. Correlations between the NPL/PTB frequency ratio without redshift correction and the chronometric levelling result are negligible. The value of the frequency ratio $R_{\mathrm{NPL,PTB}}$ measured via the EFL and using chronometric levelling is greater than the one measured using the GNSS/geoid approach by $8 \times 10^{-18}$ and lies between the two values shown in Fig. 3; its uncertainty is the same as those of the other two results.

## Data availability

The data that support the findings of this study are available from the corresponding authors on request. The results of the analysis and additional data are also available as Open Data at [URL/DOI to be inserted].

| **comparison** | **year** | **eWRMSD ($k_p$ = 2) / $10^{-18}$** | $\nu$ | **references** |
|---|---|---|---|---|
| $^{27}Al^+$ (NIST) | 2010 | 251.5 | 1 | [46] |
| $^{87}Sr$ (SYRTE) | 2011 | 1676.4 | 1 | [47] |
| $^{88}Sr^+$ (NPL) | 2013 | 125.7 | 1 | [48] |
| $^{87}Sr$ (NICT/PTB) | 2014 | 1536.7 | 1 | [49] |
| $^{171}Yb^+$ E3 (PTB/NPL) | 2014 | 1816.1 | 1 | [50] |
| $^{40}Ca^+$ (CAS) | 2015 | 447.0 | 1 | [51] |
| $^{171}Yb^+$ E3 (ITOC) | 2015 | 1945.6 | 1 | [50] |
| $^{87}Sr$ (PTB/SYRTE) | 2015 | 656.6 | 1 | [25] |
| $^{199}Hg/^{87}Sr$ | 2015 | 930.6 | 1 | [52, 53] |
| $^{87}Sr$ (ITOC) | 2015 | 833.2 | 2 | [24] |
| $^{87}Sr$ (PTB) | 2017 | 83.8 | 1 | [16] |
| $^{88}Sr/^{87}Sr$ | 2017 | 370.0 | 1 | [54, 55] |
| $^{171}Yb$ (NIST) | 2017 | 9.8 | 1 | [2] |
| $^{171}Yb^+$ E3 (PTB) | 2017 | 39.1 | 1 | [5] |
| $^{171}Yb/^{87}Sr$ | 2018 | 80.9 | 2 | [56, 57, 58, 13] |
| $^{87}Sr$ (RIKEN) | 2019 | 4.2 | 1 | [11] |
| $^{171}Yb^+$ (E3)/$^{171}Yb^+$ (E2) | 2019 | 1110.5 | 1 | [59, 60] |
| $^{115}In^+/^{87}Sr$ | 2022 | 424.8 | 1 | [61, 62] |
| $^{40}Ca^+$ (CAS) | 2023 | 97.8 | 1 | [63] |
| $^{87}Sr$ (this work, phase 1/NPL) | 2023 | 101.3 | 3 | - |
| $^{87}Sr$ (this work, phase 2/PTB) | 2023 | 89.1 | 2 | - |
| $^{176}Lu^+$ (NUS) | 2023 | 27.9 | 1 | [64] |

**Extended Data Table 1** Extended weighted RMS deviations ($k_p$ = 2) of the clock comparisons shown in Fig. 4. Comparisons are labelled by the clock species being compared and, for clocks of the same kind, the institute(s) involved or collaboration.

# Extended References

**Acknowledgements**

We thank Martina Sacher (Bundesamt für Kartographie und Geodäsie, BKG, Leipzig, Germany) for providing information regarding details on EVRF relations and uncertainties, Thomas Legero and Uwe Sterr (PTB) for providing a stable laser oscillator, Andreas Koczwara (PTB) for the development of electronics for the EFL, Max Tamussino for technical support at NPL, and Rachel Godun for discussions on the manuscript.

This measurement campaign was supported by the UK Engineering and Physical Sciences Research Council (EPSRC) Grant Ref: EP/W003279/1 for the International Clock and Oscillator Networking – ICON. RIKEN received support from the Japan Science and Technology Agency (JST) Mirai Program Grant No. JPMJMI18A1. NPL acknowledges support by the UK government Department for Science, Innovation and Technology through the UK National Measurement System programme. PTB acknowledges support by the Deutsche Forschungsgemeinschaft (DFG, German Research Foundation) under Germany's Excellence Strategy – EXC-2123 QuantumFrontiers – Project-ID 3908379.67 and SFB 1464 TerraQ – Project-ID 434617780 – within projects A04 and A05. This work was partially supported by the Max Planck–RIKEN–PTB Center for Time, Constants and Fundamental Symmetries. Optical fibre links from the French REFIMEVE network are supported by Program "Investissements d'Avenir" launched by the French Government and implemented by Agence Nationale de la Recherche with reference ANR-21-ESRE-0029 (ESR/Equipex+ T-REFIMEVE).

**Author Contributions**

I.U., Y.B.K., and M.T. contributed to operation of the RIKEN clock. C.F., I.R.H., and M.Y.H.J. contributed to operation of the NPL clock. T.L., I.N., and C.V. contributed to operation of the PTB(T) clock. S.D., J.Kl., and K.S. contributed to operation of the PTB clock. M.S. contributed to operation of the ultrastable laser at NPL. J.T. contributed to the operation of the frequency comb at NPL. E.B. contributed to operation of the frequency comb at PTB. A.A., P.B., E.C., C.C., J.-O.G., W.H., J.Kr., A.K., R.L., O.L., M.M., S.M., B.P., P.P., and M.Tø. contributed to operation of the EFL. E.B., S.D., and C.L. contributed to the evaluation of the measurements. H.D. contributed to compiling the geodetic results for NPL and PTB. S.D., I.R.H., C.L., and I.N. contributed to developing a figure of merit for clock comparisons. Y.S. contributed to the overall coordination of the consortium. H.K., I.R.H., and C.L. supervised the clock teams. A.A., C.C., J.Kr., R.L., and P.P. supervised the teams operating the EFL. All authors discussed the results and contributed to the manuscript.

**Competing Interests**

The authors declare no competing interests.

**Correspondence and requests for materials**

should be addressed to I.R.H., H.K., C.L., or Y.S.